\documentclass[preprint,12pt]{elsarticle}

\usepackage{amssymb}
\usepackage{geometry} 
\usepackage{graphics}
\usepackage{graphicx}

\usepackage{multirow} 
\usepackage{hyperref}
\usepackage{subfigure}

\journal{PSS}

\begin{document}

\begin{frontmatter}



\title{Modeling collision probability for Earth-impactor 2008~TC$_3$.}

\author{D. Oszkiewicz$^{1}$, K. Muinonen$^{1,2}$,  J. Virtanen$^{2}$, M. Granvik$^{3}$, E. Bowell$^4$}

\address{\small{ $^1$ Department of Physics, P.O. Box 64, FI-00014 University of Helsinki, Finland.} \\
\small{ $^2$ Finnish Geodetic Institute, P.O. Box 15, FI-02431 Masala, Finland.} \\
\small{ $^3$ Institute for Astronomy, 2680 Woodlawn Drive, Honolulu, HI 96822, U.S.A.} \\
\small{ $^4$ Lowell Observatory, 1400 West Mars Hill Road, Flagstaff, AZ 86001, U.S.A.} \\

}

\address{}

\begin{abstract}
We study the evolution of the Earth collision probability of asteroid 2008 TC$_3$ using a short observational arc and small numbers of observations.
To assess impact probability, we use techniques that rely on the orbital-element probability density function characterized using both Markov-chain Monte-Carlo orbital ranging and Monte-Carlo ranging. First, we evaluate the orbital uncertainties for the object from the night of discovery onwards and examine the collapse of the orbital-element distributions in time. Second, we examine the sensitivity of the results to the assumed astrometric noise. 
Each of the orbits obtained from the MCMC ranging method is propagated into the future (within chosen time bounds of the expected impact), and the collision probability is calculated as a weighted fraction of the orbits leading to a collision from the Earth. We compare the results obtained with both methods.
\end{abstract}


\end{frontmatter}


\section{Introduction}
\label{intro}
The near-Earth object 2008 TC$_3$ was a small fast rotating asteroid (discovered by R.A. Kowalski, Catalina Sky Survey), which impacted the Earth on October 7, 2008. This extraordinary object was discovered and then tracked for about 20 h before its collision. It was approaching at rather small speed of about $7.6$ km/s. Periodic light variation indicated the fast rotation of the object (period about 1.5 min) in a complex, non-principal axis ("tumbling") state \cite{SP2010}. 2008~TC$_3$ entered the atmosphere at 02:46 UTC and exploded 37 km above northern Sudan. The first prediction (S.R. Chesley, JPL) indicated explosion and a rain of small meteorites. As expected number of small fragments was found. A total of about 600 fragments (10.5 kg) were collected \cite{Nature}. The meteorites were later identified as ureilites. 
Collision probability computations and impact prediction were performed at the University of Pisa (Italy) and Jet Propulsion Laboratory (USA). Preliminary orbit computation immediately indicated collision with the Earth, and subsequent "follow-up" observations were performed, finally gathering a total of 570 astrometric observations from 26 different observatories.
NEODys (Pisa) lists five other independent possible previous close approaches of 2008~TC$_3$ with the Earth and one with Mars. Four of which have the probability approaching $100$\% and one approaching $50$\%. In Sect. 2 we describe the methods used for computing orbits and collision probability. In Sect. 3 we present our results and Sect. 4 contains conclusions.

\section{Orbits and collision probability computation}
\subsection{Orbital inverse problem}
An outline of the Bayesian treatment of the orbital inverse problem can be found, for example, in \cite{Jenni, MB93, A3, Thesis}.
The solution to the orbital inverse problem can be written as an a posteriori probability, using Bayes' theorem:

\begin{eqnarray}
p_p(\textbf{P}) = C p_{pr}(\textbf{P}) p(\psi \mid \textbf{P})& = & C p_{pr}(\textbf{P})p_{\epsilon}(\Delta \Psi  (\textbf{P})),
\end{eqnarray}
where:
\begin{itemize}
\item $p_{pr}(\textbf{P})$ is the a priori probability density function (p.d.f.), 
\item $p_{\epsilon}(\Delta \Psi (\textbf{P}))$ is the observational error p.d.f. (usually being assumed Gaussian), evaluated for the observed-minus-computed (O-C) residuals $\Delta \Psi (\textbf{P})$, 
\item $C = (\int p(\textbf{P}, \psi)d\textbf{P})^{-1}$ is a normalization constant,
\item $\textbf{P}$ denotes osculating elements $(x, y, z, \dot{x}, \dot{y}, \dot{z})^T$ or $(a,e,i,\Omega, \omega, M_0)^T$ at epoch $t_0$, 
\item $\psi$ denotes a set of astrometric observations, consisting of R.A., Dec. pairs: $\psi = (\alpha_1, \delta_1; ... ; \alpha_N, \delta_N)^T$.
\end{itemize}

To keep $p_p$ invariant in transformations from one orbital element set to another, one can use, for example, Jeffreys' a priori p.d.f. \cite{Jeff}:
	\begin{eqnarray*} 
	p_{pr}(\textbf{P}) & \propto & \sqrt{\det \Sigma^{-1}(\textbf{P})},
	\end{eqnarray*}
	
	\begin{eqnarray*} 
	\Sigma^{-1}(\textbf{P}) & = & \phi(\textbf{P})^T \Lambda^{-1} \phi(\textbf{P}),
	\end{eqnarray*}
	
where $\Sigma^{-1}$ is the information matrix (or the inverse covariance matrix) evaluated for the local orbital elements $\textbf{P}$, $\phi$ contains the partial derivatives of R.A. and Dec. with respect to the orbital elements, and $\Lambda$ is the covariance matrix for the observational errors. For details of asteroid orbital inversion theory, refer to the extensive literature available \cite{Jenni, MB93, A3, Thesis, MG}. 
To characterize the complicated a posteriori target densities, we use Monte-Carlo methods. In particular, we focus on the Markov-chain Monte-Carlo ranging technique.

\subsubsection{Markov-chain Monte-Carlo ranging}
\begin{figure}[h!]
   \includegraphics[scale=1.0]{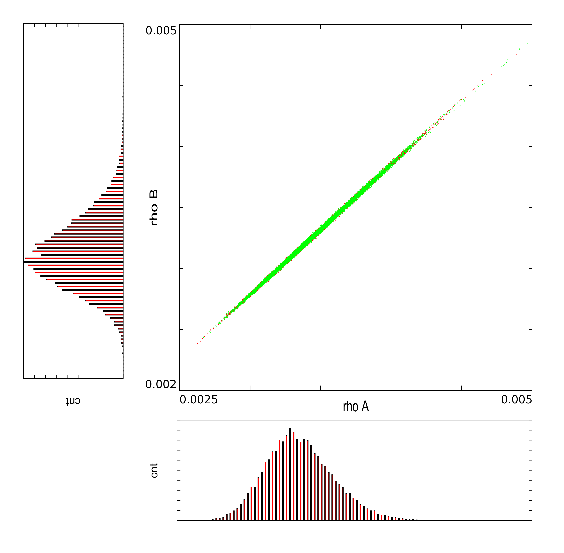} 
   \caption{Distribution of topocentric ranges $\rho_A$ and  $\rho_B$ computed using a two-body model with six observations, 0.3-arcsec astrometric noise assumption, and Jeffreys' a priori.}
   \label{ranges}
\end{figure}

Markov-chain Monte-Carlo (MCMC) ranging \cite{DO} is a nonlinear orbital sampling technique that combines the Monte-Carlo (MC) ranging method and Metropolis-Hastings (MH) algorithm. In contrast to MC ranging, in MCMC ranging the sampling is guided, with a help of arbitrary proposal density (here we use multi-normal proposal density) for inversion parameters. Subsequent orbits result in a chain (Markov chain) of samples which are not independent.
As inversion parameters the spherical coordinates $\bf{Q} = (\rho_A, \alpha_A, \delta_A, \rho_B, \alpha_B, \delta_B)$ at two chosen observation dates A and B are used. Multivariate Gaussian proposal distribution $p_t(\bf{Q}', \bf{Q_t})$ (describing the transition probability from last accepted spherical coordinates $\bf{Q_t}$ to candidate spherical coordinates $\bf{Q}'$) is adopted for candidate orbit sampling. In particular, here we use multi-normal proposal distribution with highly correlated ranges $\rho_A$ and $\rho_B$, as such a correlation is very often a case for short observational arcs \cite{DO} (also see Fig. \ref{ranges}). As the number of observations and observational time interval increases, the ranges distribution become more confined and less correlated. This simple proposal distribution in spherical coordinates phase-space corresponds to highly complicated proposal distributions in Cartesian or/and Keplerian element phase space.
To increase our chances of detection of multimodal distributions, we decided to use a number of shorter Markov chains instead of one long chain. We start our sampling with a set of random orbits (i.e., 10 sample orbits). Each of those composes a first element in a separate Markov chain. Next, a new candidate orbit is proposed (one candidate orbit for each chain) with a help of proposal densities. That orbit is then accepted into a chain or rejected according to MH acceptance criteria:

	\begin{equation}
		\begin{array}{ll}
			\rm{If \;} $a$ \geq 1, & \rm{then \;} \mathbf{P}_{t+1} = \mathbf{P}^{'}. \\
		\rm{If \;} $a$ < 1,      & \rm{then \;}  \left\{ \begin{array}{l}
			\mathbf{P}_{t+1} =\mathbf{P}^{'}, \rm{\; with \; probability \;} $a$, \\
			\mathbf{P}_{t+1} =\mathbf{P}_{t}, \rm{\; with \; probability \;} $1- a$.
			\end{array} \right.
		\end{array}
	\end{equation}		
		
	where,
	
	\begin{eqnarray}
	a &=& \frac{p_p(\mathbf{P}')}{p_p(\mathbf{P}_t)} \frac{\mid J_t \mid}{\mid J' \mid}.
	\end{eqnarray}	

$J'$ and $J_t$ are the Jacobians from topocentric coordinates to orbital parameters for the candidate and the last accepted sample respectively. $\mathbf{P}'$ and $\mathbf{P}_t$ denote a set of orbital elements, and $p_p$ is the a posteriori probability density function. 
In practice, this means that a trial orbit is always accepted if it produces a better fit to the full observational data set than the last accepted one in a given chain. The trial orbit is sometimes accepted if the fit is worse than the last accepted solution. If we have a choice between jumping to 'bad fit' and a 'very bad fit', the 'bad fit' jump is more probable. If a candidate orbit is rejected, then the last accepted solution is repeated in a chain. For more details on MCMC ranging please see \cite{DO}.
To avoid manual tuning of the proposal density covariance, we have implemented an automated version of the MCMC algorithm. In the automated version, we first run 'pre-runs' (which could be considered as a burn-in period) to the actual sampling, in order to find adequate proposal density. That proposal density is computed based on empirical samples from the pre-runs, and then updated in a manner similar to that described in \cite{AM}:

\begin{eqnarray}
p_t^{(i)}(\mathbf{Q}' \mid \mathbf{Q}_1, ... , \mathbf{Q}_n) &\sim& 
N(\mathbf{Q}_t, c_d^2 \mathbf{R}^{(i)} + c_d^2 \epsilon \mathbf{I}), 
\end{eqnarray}	
where $\mathbf{R}^{(i)}$ is the empirical $6 \times 6$ covariance matrix obtained from pre-run $i$ from $n$ samples $\mathbf{Q}_1, ... , \mathbf{Q}_n$; 
%
$c_d^2 = 2.4/\sqrt{d}$ is a scaling factor, that depends on the number of dimensions $d$ in the problem, in our case $d = 6$, $c_d \sim 0.97$. This scaling factor is optimized for for MH in the case of Gaussian targets and Gaussian proposals (For more details, please see Gelman et al. 1996 \cite{Gelman2}).  We stop updating the proposal covariance matrix after the pre-runs in order to avoid introducing biases in the sampling and to preserve the desired properties of Markov chains.
The number of pre-runs depends on, for example, convergence diagnostic checks. The minimum number of pre-runs is two. The first proposal covariance matrix has to be user defined. Then, we start the automated MCMC ranging with the first pre-run and require 500 sample orbits. From that first pre-run, we compute the new proposal covariance matrix and continue with a second pre-run (1000 required orbits). A third pre-run is possible if the final number of required orbits is 50000 or more.
After those pre-runs, we continue with iterative runs until all of the below conditions are fulfilled: 
\begin{itemize}
\item the algorithm has decided that it has reached the stationary distribution (please see the next section for more details),
\item the jump ratio in the last run is between 15-50\% (recommended MCMC ratio),
\item	 the difference in the obtained maximum likelihood p.d.f in subsequent runs is smaller than a threshold value ($\mid 2.0 \log{pdf_i/pdf_{i-1}} \mid < 2.0$).
\item the difference in the obtained empirical covariance matrix in sub-sequentive runs is smaller than a threshold value ($\mid 2.0 \log{R_i/R_{i-1}} \mid < 2.0$).
\end{itemize}

\begin{figure}[h!]
   \includegraphics[scale=1.0]{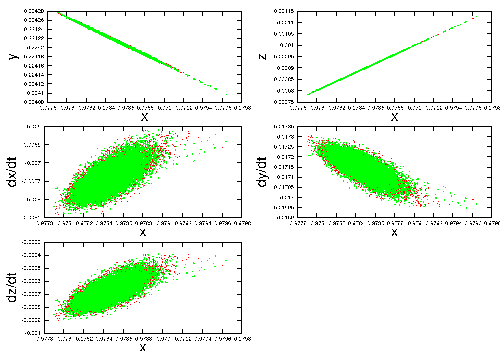} 
   \caption{Distribution of orbital elements (here: Cartesian ecliptic), computed using two-body model with six observations, 0.3-arcsec astrometric noise assumption and Jeffreys' a priori.}
   \label{orbel}
\end{figure}

\begin{figure}[h!]
   \includegraphics[scale=1.0]{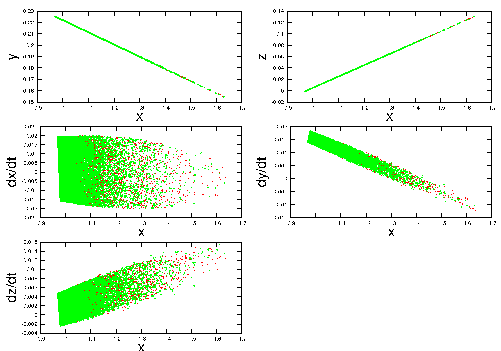} 
   \caption{MCMC ranging with Jeffreys' a priori concentrates on the orbits with smaller Cartesian ecliptic coordinate $x$.}
   \label{Jeff}
\end{figure}

\begin{figure}[h!]
   \includegraphics[scale=1.0]{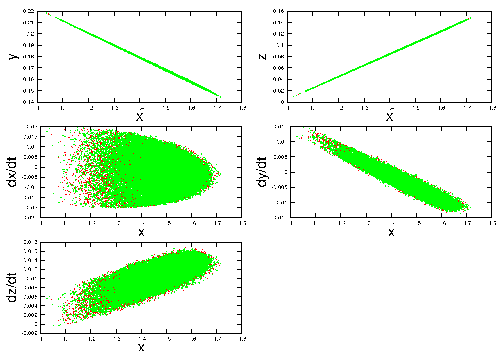} 
   \caption{MCMC ranging with uniform a priori concentrates on the orbits with larger Cartesian ecliptic coordinate $x$.}
   \label{Unif}
\end{figure}
Even though the conditions are designed to ensure that we have reached the stationary distribution, one should always remember that convergence might still not have been reached (see the next section). If those conditions were not fulfilled and the number of iterations exceeded the maximum number of iterations (here we use five iterations after the pre-runs), the program exits with a warning that the stationary distribution has not been reached. Example results of a converged run are shown in Fig. \ref{orbel}.

\subsubsection{MCMC ranging convergence diagnostics}
\begin{figure}[htbp]
   \centering
   \includegraphics[width=\textwidth]{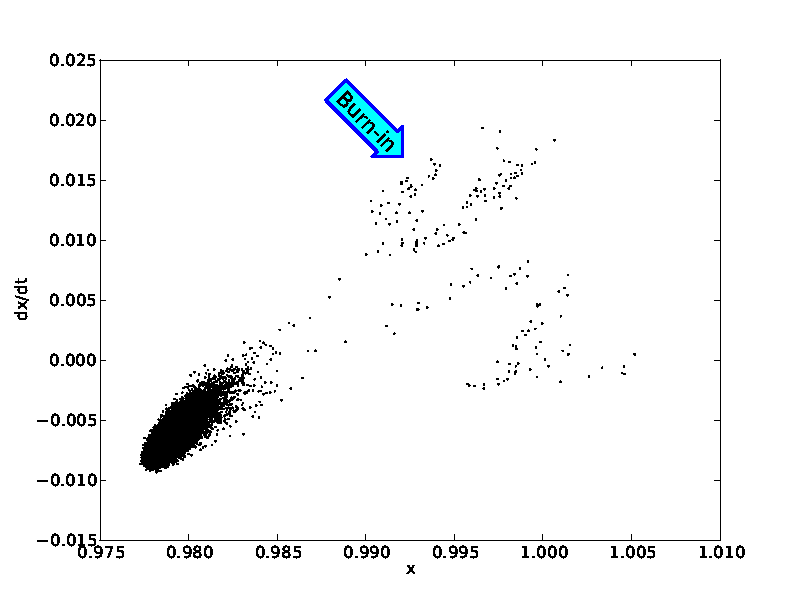} 
   \caption{The marginal distribution of ecliptic coordinates $\dot{x}$ and $x$, computed for asteroid 2008 TC$_3$ using five observations and 0.5-arcsec astrometric noise assumption. The arrow indicates the burn-in samples, before reaching the stationary distribution.}
   \label{burnin}
\end{figure}

After obtaining sets of possible orbit solutions for each chain, we perform convergence diagnostics, which is a critical part of MCMC algorithms. Convergence diagnostics relate to the idea that a Markov chain, after a sufficient number of iterations, will eventually converge to the stationary distribution (target distribution), starting from any point in the phase space and then mixing in that part of phase space forever. The proof of convergence can be found for example in Roberts and Rosenthal 2004 \cite{GOR}. The main concerns are: (1) convergence to the stationary distribution; (2) the length of the burn-in period (for example, see Fig. \ref{burnin}); and, finally, (3) number of iterations it will take to summarize the posterior distribution \cite{cowles1996markov}. 

In general, there are two approaches to monitoring the convergence: monitoring of one chain for a long time or monitoring many chains for shorter periods of time. At present, a set of convergence diagnostics methods is available for MCMC algorithms for both of these approaches. The main idea usually involves running the Markov chain repeatedly from different initial states and checking if the chains all converge to approximately the same distribution or, in the case of a single chain, running it for long time and checking if it "looks stable". The diagnostics help identify problems with convergence, but cannot prove that the convergence has occurred.

\begin{figure}[htbp]
   \centering
   \includegraphics[width=\textwidth]{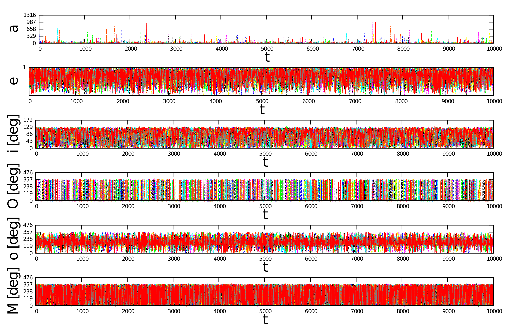} 
   \caption{A history plot generated from MCMC sampling of orbits for asteroid 2008 TC$_3$, assuming 1.5-arcsec astrometric noise and including four observations for a time interval of 0.003 day.}
   \label{history}
\end{figure}

The basic convergence testing methods include history plots (plot of sampled parameters against the number of iterations, also called trace plots) and marginal distribution plots. If the algorithm converged to a stable distribution, the history plot will move around the mode of the distribution, and all the chains will cover the similar parameter range. Trending in trace plots can indicate lack of convergence. Figure \ref{history} shows an example trace plot. If the separate chains converged to different areas in the phase space, we may be able also to see "lumpy" posterior marginal distributions, with distinct solution clouds. This effect can be sometimes also seen in the observed-minus-computed (O-C) residuals plots. To test the randomness of a data set, autocorrelation function plots (correlograms) are often used. Correlograms involve computing the correlation between data sets from a single chain and separated by some lag. If random, all the correlations will be near zero.
Gelman--Rubin \cite{Gelman} proposed a statistical indicator, a so-called "shrink factor" R, involving the average within-chain variances, and the variance between the means obtained from $m$ parallel over-dispersed chains. 
Once convergence is reached, the between chains and within chains variance should be almost equivalent because variation within the chains and variations between the chains should coincide, so R should approximately equal one.
Rafery and Lewis \cite{raf} proposed a method for estimating the total numbers of iterations needed, the length of the burn-in period based on chosen quantile, required accuracy, required probability to obtain specified accuracy and convergence tolerance. Gewke \cite{gewke} indicated the use of methods from spectral analysis in testing for convergence. The method is graphical and based on running means of parameters and can be used in, for example, checking for the length of the burn-in period.
Zellner and Min \cite{zellner} proposed a criterion based on conditional probability to test for convergence to a correct distribution.
Extensive review of MCMC convergence diagnostics methods can be found in for example Mengersen et al. 1999 and Cowles and Carlin 1996 \cite{diag, diag1}. Topic of biases induced in the MCMC diagnostics is considered in for example Clowes et al. 1999 \cite{clowes}.
In our analysis, we make use of trace plots, kernel plots, O-C residuals plots, and Gelman--Rubin diagnostics.

\subsection{Collision probability}
To assess the impact probabilities, we use techniques that rely on the orbital-element probability density function (p.d.f.) characterized using MCMC orbital ranging \cite{DO, MG} and MC ranging \cite{Jenni}. After obtaining a set of possible orbit solutions (here we use $5 \times 10^4$ different MCMC orbits for all the objects or/and $5 \times 10^4$ MC orbits), we propagate each of the orbits into the future to check for a possible collision with the Earth within the determined impact interval $\tau$ (time interval when collisions are possible).
We step through this interval depending on the approximated time needed for the object to hit the Earth, given as it would be heading straightforward to the Earth. The next propagation step is therefore fixed as the time interval needed for the object to reach the Earth, given it would be heading head-on with our planet and depending on the current velocity and distance. The collision probability is given by \cite{JVKM, KM}:
\begin{equation}
P_c (\tau) = \int_0^{R} d\rho \hspace{0.2 cm} p_p(\rho, \tau),
\end{equation}
where $R$ is the radius of the planetary body and $p_p(\rho, \tau)$ is defined as:
\begin{equation}
p_p(\rho', \tau) = \int d\textbf{P} p_p(\textbf{P})\delta_D(\rho'-\rho(\textbf{P}, \tau)),
\end{equation}
where $\rho(\textbf{P}, \tau)$ is the minimum distance of the small body from the center of the large (planetary) body, $\delta_D$ is Dirac's function, $\textbf{P}$ denotes orbital elements, $p_p$ is the a posteriori p.d.f. of the orbital elements leading to a collision. For more details, please see Virtanen and Muinonen 2006 and Muinonen et al. 2001 \cite{JVKM, KM}. In case of 2008 TC$_3$, we used an impact interval of Oct. 6-8, 2008. In the case of MC results, the orbits are weighted with the a posteriori p.d.f.; in the case of MCMC results, the weight of an orbit equals the number of repetitions of the orbit.

\section{Results and discussion}
To study the evolution of collision probabilities for the asteroid 2008 TC$_3$ based on a short observation arc, we use a limited number of observations from the discovery night only. Here we use the first six observations obtained on October 6, 2008. We divide this observational data into separate small sets. We start with two observations and then the subsequent sets contain one more observation than the previous one. We end up with five sets of different length (please see Table \ref{obs} for details).

\begin{table}[htbp]
   \centering
      \begin{tabular}{| l | l | l | l | l | l |}  \hline
      Number of Observations & 2 & 3 & 4 & 5 & 6 \\ \hline
      Time interval [days] & 0.01 & 0.02 & 0.03 & 0.04 & 0.06 \\ \hline
       \end{tabular}
   \caption{Observational data sets.}
   \label{obs}
\end{table}

First, we obtain a set of MCMC and MC orbits using each of those sets, and then we perform the collision probability computation.
We repeat this process for different assumptions of the observational noise and different dynamical models. The results obtained are listed in Table \ref{coll} for the two-body case and in Table \ref{coll0} for the n-body case.

\begin{table}[h!]
 \centering
\begin{tabular}{|l|l|l|l|}
\hline
Nr. of & Noise & MC & MCMC \\
obs. & [arcsec]: & ranging: & ranging:  \\ 
& &Jeffrey's apriori &Jeffrey's apriori  \\ \hline
\multirow{5}{*}{2} 
 & 0.3 &  0.75 (1536) & 0.58 (12587)  \\ 
 & 0.5 & 0.62 (1722) & 0.57 (12157)  \\ 
 & 1.0 & 0.65 (2731) & 0.61 (10815)  \\ 
 & 1.5 & 0.67 (2586) & 0.61 (11136)   \\ 
 & 2.0 & 0.69 (2274) & 0.62 (11124)  \\ \hline
\multirow{5}{*}{3} 
 & 0.3 & 0.008 (178) & 0.008 (308)  \\
 & 0.5 & 0.02 (282) & 0.01 (490)    \\
 & 1.0 & 0.14 (663) & 0.11 (1218) \\
 & 1.5 & 0.54 (1403) & 0.22 (3062) \\ 
 & 2.0 & 0.39 (3190) & 0.28 (4521) \\ \hline
\multirow{5}{*}{4} 
 & 0.3 & 0.0007 (191) & 0.0006 (83)  \\
 & 0.5 & 0.005 (264) & 0.004 (308)  \\
 & 1.0 & 0.02 (388) & 0.01 (725)   \\
 & 1.5 & 0.04 (591) & 0.02 (1042) \\ 
 & 2.0 & 0.04 (967) & 0.06 (1666) \\ \hline
 \multirow{5}{*}{5} 
 & 0.3 & 0 (0) & 0 (0) \\
 & 0.5 & 0.0002 (325) & 0.0001 (11) \\
 & 1.0 & 0.007 (414) & 0.005 (457) \\
 & 1.5 & 0.007 (516) & 0.01 (739) \\ 
 & 2.0 & 0.02 (566) & 0.02 (1065) \\ \hline
  \multirow{5}{*}{6} 
 & 0.3 & 0 (0) & 0 (0) \\
 & 0.5 & 0.0000008 (2) & 0 (0) \\
 & 1.0 & 0.0005 (449) & 0.0006 (63)\\
 & 1.5 & 0.004 (490) & 0.004 (417) \\ 
 & 2.0 & 0.01 (568) & 0.009 (694) \\ \hline
 \end{tabular}
\caption{Collision probability, two-body approach. Number of different collision orbits (out of 50 000 different orbit solutions) given in brackets.}
\label{coll}
\end{table}

\begin{figure}[htbp]
   \centering
   \includegraphics[width=\textwidth]{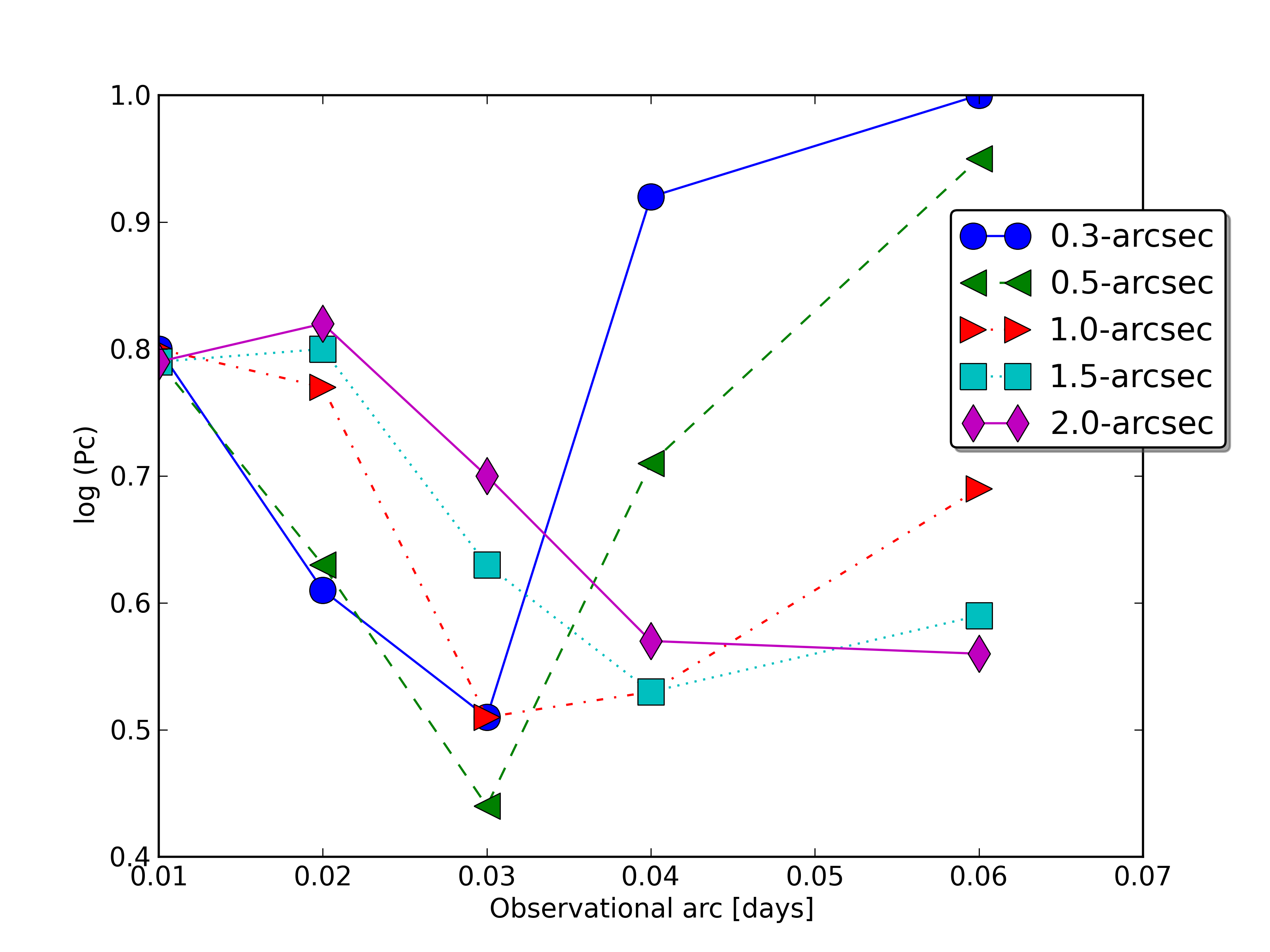} 
   \caption{Time evolution of collision probabilities for 2008 TC$_3$ using MCMC ranging generated orbits with Jeffreys' a priori (n-body approach). }
   \label{mc1}
\end{figure}

\begin{table}
 \centering
\begin{tabular}{|l|l|l|}
\hline
Nr. of & Noise & MCMC \\
obs. & [arcsec]: & ranging: \\ 
& &Jeffrey's apriori   \\ \hline
\multirow{5}{*}{2} 
 & 0.3 & 0.80 (15281)   \\ 
 & 0.5 & 0.79 (15114)   \\ 
 & 1.0 & 0.80 (14584)  \\ 
 & 1.5 & 0.79 (13991)  \\ 
 & 2.0 & 0.79 (13585) \\ \hline
\multirow{5}{*}{3} 
 & 0.3 & 0.61 (2813)  \\
 & 0.5 & 0.63 (3271) \\
 & 1.0 & 0.77 (6025)  \\
 & 1.5 & 0.80 (8412)  \\ 
 & 2.0 & 0.82 (12113)  \\ \hline
\multirow{5}{*}{4} 
 & 0.3 & 0.51 (14960)  \\
 & 0.5 & 0.44 (7878) \\
 & 1.0 & 0.51 (4742) \\
 & 1.5 & 0.63 (6116)  \\ 
 & 2.0 & 0.70 (8678)  \\ \hline
 \multirow{5}{*}{5} 
 & 0.3 & 0.92 (44053)  \\
 & 0.5 & 0.71 (28966)   \\
 & 1.0 & 0.53 (12620)   \\
 & 1.5 & 0.53 (9611) \\ 
 & 2.0 & 0.57 (8930)    \\ \hline
  \multirow{5}{*}{6} 
 & 0.3 & $\approx$ 1.0 (49771)   \\
 & 0.5 & 0.95 (46079)   \\
 & 1.0 & 0.69 (27786)  \\
 & 1.5 & 0.59 (16348)  \\ 
 & 2.0 &  0.56 (11939) \\ \hline
 \end{tabular}
\caption{Collision probability (as a weighted fraction of collision orbits), n-body approach. Number of different collision orbits (out of 50 000 different orbit solutions) given in brackets.}
\label{coll0}
\end{table}

\begin{figure}[p]
  \centering
	    \subfigure[ Green - the marginal distribution of the orbital elements obtained for asteroid 2008 TC$_3$ using five observations and 0.5-arcsec noise. Blue - orbits leading to a collision with the Earth from the green set. Red - orbits leading to a collision with Earth obtained using 3 observations and 2.0-arcsec noise. ]{
      \label{TorroColl}
      \includegraphics[scale=0.45]{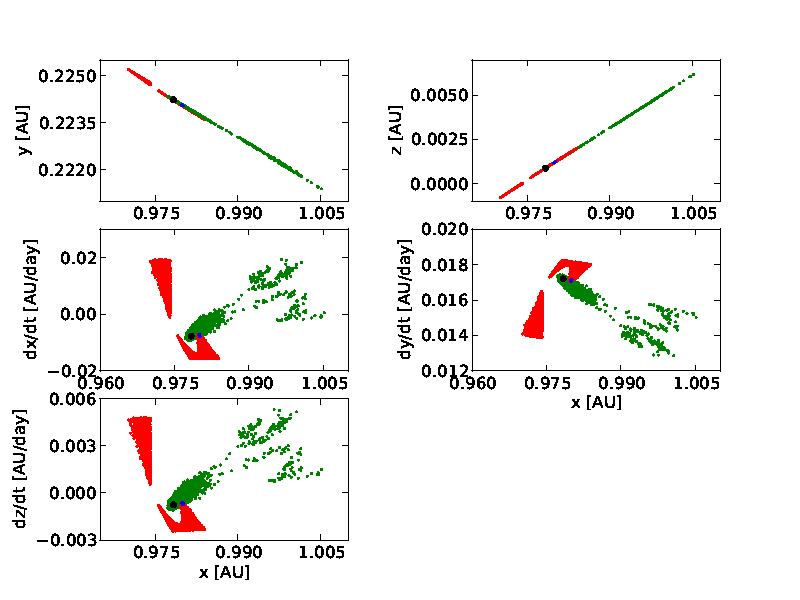}
    }
    \subfigure[ Green - the marginal distribution of the orbital elements obtained for asteroid 2008 TC$_3$ using 6 observations and 0.3 arcsec noise assumption. Red - orbits leading to a collision with the Earth obtained using 3 observations and 2.0-arcsec noise.  ]{
      \includegraphics[scale=0.45]{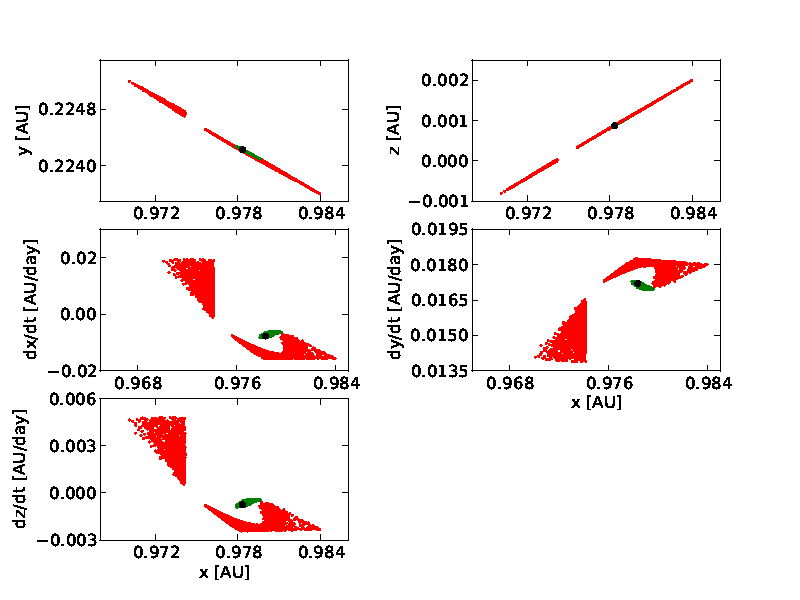}
    }
    \caption{Shift to non-impact zone in the phase space in the two-body approximation (adding the 6th observation leads to a vanishing collision probability). The black dot indicates the least-squares solution obtained with the full observational data set (see table \ref{ls}). This orbit does not lead to a collision with the Earth in the specified time interval when propagated using the two-body model.
        (For interpretation of the references to color in this figure caption, the reader is referred to the web version of this article.)}
    \label{collOrb1}  
\end{figure}

\begin{figure}[p]
  \centering
	    \subfigure[ Green - the marginal distribution of the orbital elements obtained for asteroid 2008 TC$_3$ using five observations and 0.5-arcsec noise. Blue - orbits leading to a collision with the Earth from the green set. Red - orbits leading to a collision with the Earth obtained using 3 observations and 2.0-arcsec noise. ]{
      \label{TorroColl}
      \includegraphics[scale=0.45]{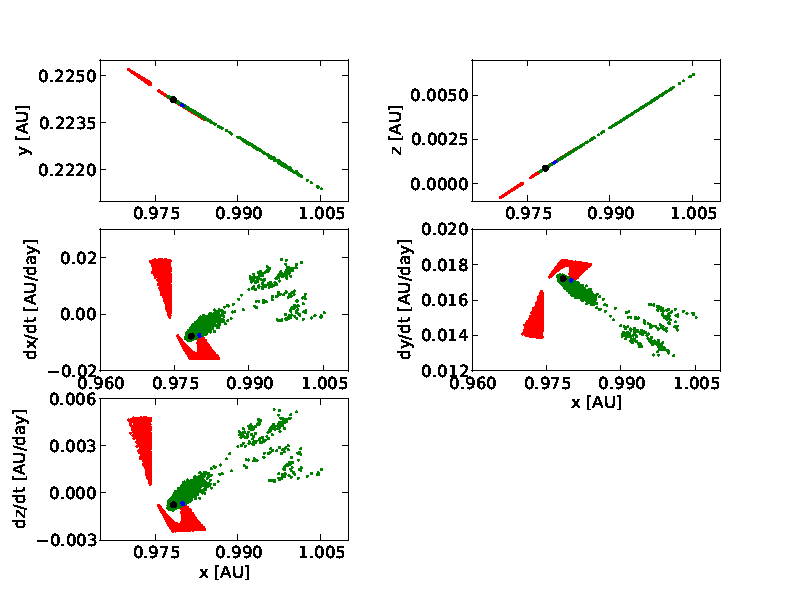}
    }
    \subfigure[ Green - the marginal distribution of the orbital elements obtained for asteroid 2008 TC$_3$ using 6 observations and 0.3-arcsec noise. Blue - orbits leading to a collision with the Earth from the green set. Red - orbits leading to a collision with the Earth obtained using 3 observations and 2.0-arcsec noise. ]{
      \includegraphics[scale=0.45]{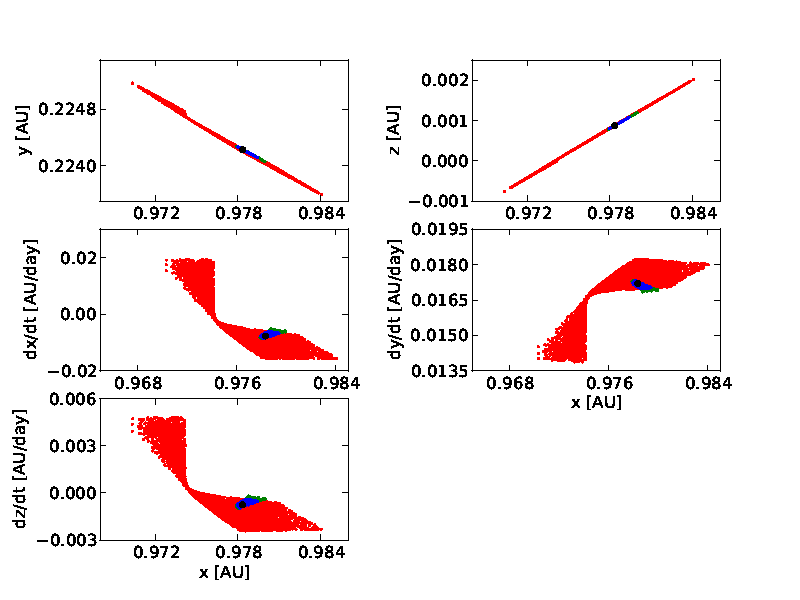}
    }
    \caption{Shift to impact phase space in the n-body approach. The black dot indicates the least-squares solution obtained with the full observational data set (see table \ref{ls}).
    (For interpretation of the references to color in this figure caption, the reader is referred to the web version of this article.)}
    \label{collOrb2}  
\end{figure}

We list the collision probability computed for the two-body approximation for MC and MCMC ranging in Table \ref{coll} for comparison between the two methods.
The results obtained from both methods agree very well. 
In Table \ref{coll0}, we list the impact probability computed in the n-body approximation, for different a priori p.d.f. assumed. The collision probability computed with the n-body and Jeffreys a priori results in a high collision probability already starting with 2 observations, and continues to grow with subsequent observations added reaching approximately $1.0$ for 6 observations and 0.3 arcsec astrometric noise. Adding additional observations (above 6 observations) to the data sets used leads to a collision probability approaching $1.0$. Collision orbits are located in a small phase-space area, as compared to the phase space of all permitted orbital solutions, and have relatively high weight factors. 
The usage of Jeffreys' a priori shifts the solutions area to concentrate more on the orbits with smaller semi-major axes, whereas the usage of a uniform a priori causes a shift towards larger semi-major axes orbits (please see Figs. \ref{Jeff} and \ref{Unif} for comparison of marginal distributions for orbital elements using Jeffreys' and uniform a priori).  Results should be tested against different a priori p.d.f. assumptions.

In the case of short observational arcs, the topocentric ranges at the observation dates A and B are very often highly correlated. An example of this can be found in Fig. \ref{ranges}, which presents the distribution of topocentric ranges $\rho_A$ and  $\rho_B$ computed using a two-body model, six observations, 0.3 arcsec astrometric noise, and Jeffreys a priori.
The usage of a two-body dynamical model led to a shift of solutions to a non-collision phase space (see Fig. \ref{collOrb1}), whereas the usage of the n-body model led to a shift into the collision phase space (see Fig. \ref{collOrb2}). For comparison, we also computed a least-squares solution, please see Table \ref{ls}.

Both methods show a sensitivity to the noise assumption. In general, the sensitivity to the noise should become small for extensive observational arcs (the data contains more information). However, in our case (maximum six observations and 0.006-day time interval), such a stability cannot be yet observed. The sensitivity of the derived probabilities to the noise assumption can be estimated as \cite{KM, JVKM} : 
\begin{equation}
Sensitivity \approx \mid \frac{\Delta \log P_c}{\Delta \sigma} \mid .
\end{equation}
Large nonzero values indicate that the results are highly sensitive to the noise assumption. In Tables \ref{sen1} and \ref{sen2}, we list the sensitivity estimation (per arc sec) for collision computation for all the sets, for both methods, using the two-body and n-body approaches. 

\begin{table}[htbp]
   \centering
   \begin{tabular}{|c|c|c|}
   \hline
      Obs. arc [days] & MC 2-body & MCMC 2-body  \\ 
      & Jeffreys apriori & Jeffreys apriori  \\ \hline
      0.01 & 0.02 &  0.02  \\
      0.02 & 1.8 &  1.0 \\
      0.03 & 0.99 &  0.7 \\
      0.04 & 1.33 &  1.53  \\
      0.06 & 2.73 &  1.18 \\ \hline
   \end{tabular}
   \caption{Sensitivity to the noise assumption for MC and MCMC ranging techniques (2-body approach).}
   \label{sen1}
\end{table}

\begin{table}[htbp]
   \centering
   \begin{tabular}{|c|c|}
   \hline
      Obs. arc [days] & MMCC n-body  \\ 
      & Jeffreys apriori  \\ \hline
      0.01& 0.0032  \\
      0.02 & 0.08 \\
      0.03 & 0.13  \\
      0.04 & 0.12  \\
      0.06 & 0.15  \\ \hline
   \end{tabular}
   \caption{Sensitivity to the noise assumption for MC and MCMC ranging techniques (n-body approach).}
   \label{sen2}
\end{table}


We indicate the need for more reliable methods of collision probability estimation (less sensitive to the noise assumption) for poorly observed objects. Here we also perform collision probability computation using 
MCMC computed orbits, where astrometric noise has been included in the list of sampling parameters. Table \ref{coll4} shows a result of that exercise. The time evolution of this probability differs from the one computed using a fixed assumed astrometric error. The impact probability also starts with a quite high value, then it grows, drops down, stabilizes and grows again (for data sets exceeding six observations). We have assumed that all the observations have the same astrometric uncertainties.
All the observations that we used come from a single survey, that is the Mt. Lemmon Survey (1.5-m telescope). We have selected a priori information on the astrometric uncertainties at Mt. Lemmon based on astrometric residuals of all objects observed in 2008 at that observatory. We have assumed a Gaussian a priori for the noise with a covariance matrix computed from those residuals, and mean of 0.3 arc seconds. One could go even further by sampling the astrometric accuracy for each observation separately, for each observatory or for each observer, and consequently using different a priori information for those. This kind of analysis is, however, beyond the scope of this paper.

\begin{table}
\centering
\begin{tabular}{|l|l|}
\hline
Nr. of & MCMC \\
obs. & ranging:  \\ \hline
2 & 0 (0) \\
3 & 0.43 (43491)\\
4 & 0.70 (55407)\\
5 & 0.55 (35895) \\
6 & 0.54 (35829)  \\ \hline
 \end{tabular}
\caption{Collision probability (as a weighted fraction of collision orbits) computed with astrometric noise sampling, n-body dynamical model, and Jeffrey's apriori used. Number of different collision orbits (out of 50 000 different orbit solutions) given in brackets.}
\label{coll4}
\end{table}

\begin{table}[htbp]
   \centering
   \begin{tabular}{|c|c|c|c|c|}
   \hline
Element 	& Value 		& 1-$\sigma$ variation & Units \\ \hline
$x$ 		& 0.978354962      & 0.57E-07			& AU \\
$y$ 		& 0.2242293386    & 0.45E-08 		& AU \\
$z$		& 0.871659598E-03 &0.11E-07	& AU \\
$\dot{x}$ & -0.776631371E-02  & 0.55E-07 	& AU/day \\
$\dot{y}$ &  0.1720023476E-01 & 0.45E-08	& AU/day \\
$\dot{z}$ & -0.755199990E-03 & 0.11E-07	& AU/day \\  \hline
   \end{tabular}
   \caption{Least square solution (cartesian ecliptic coordinates) computed with the complete set of observations 859 (70 discarded) for the epoch of 2008/10/6.0 using OpenOrb software \cite{MG}. 1.0 arcsec astrometric noise assumption and n-body dynamical model.}
   \label{ls}
\end{table}

\section{Conclusions}

We have computed and examined collision probabilities using the MC and MCMC ranging methods for the Earth-impactor asteroid 2008 TC$_3$ using small amounts of data. 
We have considered sampling of the astrometric uncertainty as an additional parameter to the orbital elements, which can be important for cases where that uncertainty is not known or cannot be assumed. 
We indicate the need for more reliable methods of collision probability estimation (less sensitive to the noise assumption) for poorly observed objects.
New ground-based and satellite sky surveys (like the forthcoming Gaia survey), due to their improved accuracy, will significantly improve the computation of collision probabilities based on small amounts of data.
NEO impact monitoring is currently not included in the Gaia processing pipeline and will have to be done outside the main data processing. Fast access to the Gaia data is essential for objects such as the Earth impactor 2008 TC$_3$. Earlier data realizes for NEOs should be therefore considered.

\section{Acknowledgments}
Research by DO supported, by EC Contract No. MRTN-CT-2006-033481 (ELSA).
In this research we made use of the OpenOrb - open source orbit computation software \cite{MG}.


\end{document}